# Phase transitions in geometrothermodynamic model of charged generalized-NUT black holes

Halina Grushevskaya[1] and Nina Krylova[2]

*Belarusian State University, Minsk, Belarus,*
*[1]grushevskaja@bsu.by, [2]krylovang@bsu.by*

**Abstract**. Modern cosmological models are constructed in the framework of thermodynamic approaches developed within a Van der Waals – Maxwell theory of the first-order phase transitions. In the present work we study a geometrothermodynamics of two-dimensional first-order phase transition with the distribution of relaxation times in a configuration space which describes a spacetime with Newman–Unti–Tamburino-like metric. We utilized the geometrothermodynamical approach to construct the model of a charged generalized-NUT black hole. We reveal following features of the black-hole phase transition: there are series of bifurcations of pitchfork type in dependences of the Gibbs free energy on the Hawking temperature, and although a scalar Berwald curvature of space changes sign in the phase transition, black-hole stability depends on sign of the curvature after the transition.

## Introduction

Recently, series of modern cosmological models are constructed in the framework of thermodynamic approaches, which are developed on the grounds of standard thermodynamics of the 1-st order phase transitions. In particular, there are examined the large class of asymptotically anti-de-Sitter black holes [1], charged de Sitter black holes [2], black hole in dark matter background [3]. But, the cosmological models which are based on Van der Waals – Maxwell theory of the liquid-gas first-order phase transitions, should meet a demand of the agreement of different thermodynamic representations (for example, between energy (black-hole mass $M$) and Gibbs free energy $G$ representations) to provide comprehensive description of the model. However, this demand extremely complicates the calculations because of that a thermodynamic geometry of the cosmological model has been constructed without taking into account a cosmological constant $\Lambda$ or scalar fields [1]. To describe a model of the universe filled by barotropic fluid (dark energy or a scalar field named quintessence) with the Weyl's canonical coordinates and negative cosmological constant $\Lambda$, the static solutions for axially symmetric metric for black holes in vacuum have been acquired [4,5]. Axially symmetric Newman–Unti–Tamburino (NUT) metrics pretend to play role of such metrics. Although some their thermodynamic properties have been studied in $f(R)$ gravity [6] and within Rastall theory of gravity [7], due to a complexity of the NUT-models a thermodynamic geometry has not been constructed yet. It has been proposed in [8] that the features of behaviour of galaxy rotation curves can be considered as effects of velocity on matter. Velocity effects are typical for a two-dimensional (2D) first-order phase transition with the distribution of relaxation times that can be described in a configuration space $(\vec{r}, \xi, \dot{\vec{r}}_\xi, \dot{\xi}_s, s)$ possessing the properties of contact manifolds which are invariant with respect to Legendre transformations [9]. Here $\vec{r}$ is a 2D radius-vector, $\xi$ is a time-like coordinate, dots are defined the derivatives with respect to an evolution parameter $s$. This configuration space present a thermodynamic phase space $\mathcal{T} = \left\{\vec{r}, \vec{r}_\xi, \dfrac{ds}{d\xi}\right\}$ as the dynamical-system phase-space $\{\vec{r}, \vec{r}_\xi\}$, which is augmented by the subspace spanned on the tangent vector $\dfrac{ds}{d\xi}$. The bundle $\mathcal{T}$ includes the four-dimensional (4D) space of the two pairs $\{\vec{r}, \vec{r}_\xi\}$ and admits a non-trivial connection determined by a Reeb vector field $\dfrac{\partial}{\partial s}$ of the contact manifold (see [10-12] and references therein). Therefore, $\mathcal{T}$ is a five-dimensional (5D) contact manifold. Because the 1st-order phase transition proceeds on the interface, the thermodynamic metric is an axially symmetric one. In the present work we utilized the geometrothermodynamical approach developed in [13] to the first-order phase transition of charged anti-de-Sitter black holes to construct the model of an electrical charged and dyonic charged generalized-NUT black hole.

The goal of the paper is to construct and to study the geometrothermodynamics of 2D 1-st order phase transition on the 5D contact statistical manifold referred to a spacetime with generalized Newman–Unti–Tamburino (NUT) metric holding a functional NUT-parameter. With the goal we approximate the numerically calculated dependences of relaxation times by lapse functions of different types with constant NUT-parameters and then analyze corresponding dependences of the Gibbs free energy on the Hawking temperature. Our finding is that a cascade of bifurcations of pitchfork type is a feature of the phase transition in the electrically charged and dyonic black NUT-holes.

### A theoretical formalism

A generalized NUT-metric with a functional NUT-parameter $n(\theta)$ for 4D spacetime reads [13]

$$dl^2 = f(R)\left[ dx^0 - \frac{45i}{24\pi}\left(\cot\frac{\theta}{2} - i\right)^2 \sin^2\frac{\theta}{2} d\varphi/(2\pi)\right]^2 - \frac{\left[dR^2 + R^2\left(d\theta^2 + \sin^2\theta d\dot\phi^2\right)\right]}{f(R)}, \quad (1)$$

where the NUT-parameter depends on the angle $\theta$ and becomes an imaginary constant at $\theta \to \pi, 0$:

$$n(\theta)\sin^2\frac{\theta}{2}\bigg|_{\theta\to\pi,0} \equiv -\frac{45i}{192\pi^2}\sin^2\frac{\theta}{2}\left(\cot\frac{\theta}{2} - i\right)^2\bigg|_{\theta\to\pi,0} = \pm\frac{45i}{192\pi^2}, \quad (2)$$

$f(R)$ is a lapse function, $(R,\theta,\phi)$ are spherical coordinates, $x^0$ is a time coordinate. In an entropy representation the metric (1) is produced from the thermodynamic metric [13]

$$dl^2 = L(\vec{r},\dot{\vec{r}},\dot\xi)ds^2 = -\tilde{p}Vr^5 e^{\frac{2V\xi}{r}}\frac{\dot\xi}{\dot r}d\xi\, ds + \tilde{U}(\xi,r)\dot\xi\, ds^2 + m\frac{(\dot r^2 + r^2\dot\phi^2)}{2\dot\xi}ds^2 \quad (3)$$

by a double Wick rotation of the form $\xi = -i\tau$ and $s = -ix^0$ and relations $dR = \dot r dx^0$, $d\phi = \dot\phi dx^0$, and $R \equiv r$. Here $V$ is a modulus of compression (or expansion) rate; $\tilde{p}$, $c$ and $m$ are model parameters, a function $\tilde{U}(\xi,r)$ named reduced potential function is expressed through a potential function $U(\xi,r)$ for a 1st-order 2D phase transition as $\tilde{U}(\xi,r) = U(\xi,r) - \tilde{p}Vr^5 e^{\frac{2V\xi}{r}}/\dot r - mc^2$. The lapse function $f(r)$ is determined by $\dot\xi$ as $\dot\xi = -f(r)m/2$. In this paper an analysis is performed utilizing the reduced potential function $\tilde{U} \equiv \tilde{U}_1$ and its approximation $\tilde{U}_2$ in a following explicit form:

$$\tilde{U}_1 = \tilde{p}V^2\xi\left(e^{\frac{2\xi V}{r}}\left(\frac{1}{2}r^4 + \frac{1}{3}r^3\xi V + \frac{1}{3}r^2\xi^2 V^2 + \frac{2}{3}r\xi^3 V^3\right) - \frac{4}{3}\xi^4 V^4 \mathrm{Ei}\left(\frac{2\xi V}{r}\right)\right)\frac{\dot\xi}{\dot r} - mc^2$$
$$-\tilde{p}\left(e^{\frac{2\xi V}{r}}\left(-\frac{3r^5}{4} + r^4\xi V + \frac{3}{4}r^3\xi^2 V^2 + \frac{5}{6}r^2\xi^3 V^3 + \frac{11}{6}r\xi^4 V^4 - \frac{\xi^5 V^5}{3}\right) - \frac{2}{3}\xi^5 V^5\left(6 - \frac{\xi V}{r}\right)\mathrm{Ei}\left(\frac{2\xi V}{r}\right)\right), \quad (5)$$

$$\tilde{U}_2 = \tilde{p}\left\{\left[-\frac{4}{3}r^5 + \frac{16}{15}(V\xi)r^4 + \frac{1}{30}(V\xi)^2 r^3 + \frac{1}{45}(V\xi)^3 r^2\right.\right.$$
$$\left.\left. + \frac{1}{45}(V\xi)^4 r + \frac{2}{45}(V\xi)^5\right]e^{\frac{2V|\xi|}{r}} + \frac{4}{45}\left((V\xi)^5 - \frac{(V\xi)^6}{r}\right)\text{Ei}\left(\frac{2V\xi}{r}\right)\right\} - mc^2. \quad (6)$$

The thermodynamic metric (3) is implemented on a contact statistical manifold for the first-order 2D phase transition. In the fig.1 the typical dependences of potential functions $U_1$ and $U_2$ on radius $r$ are shown. As one can see, the approximate function $U_2$ has not so sharp local minimum as the function $U_1$.

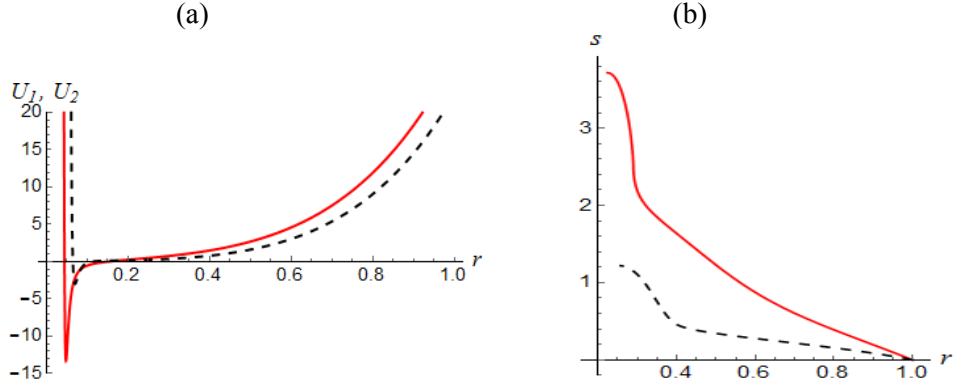

Fig 1. (a) The typical dependences of potential functions $U_1$ (solid) and $U_2$ (dashed) on radius $r$ at following values of parameters: $\dot{r}/\dot{\xi} = 0.027$, $V=0.224$, $\xi = 2$, $\tilde{p} = 1$. (b) The isotherms $s(r)$ (geodesics) obtained by numerical calculations. Thermodynamic metric function parameters and potential functions used: $\tilde{p} = 1$, $mc^2 = 1$, $m = 10^{-5}$, $U_1$ (solid); (b) $\tilde{p} = 1$, $m = 1$, $c = 1$, $U_2$ (dashed).

**A black-hole thermodynamic geometry analysis**

To study thermodynamic properties of the metric (1) using the thermodynamic space with the metric (2) we transform the latter to a following auxiliary Finsler metric function:

$$F_i^2 = -L(\vec{r},\dot{\vec{r}},\dot{\xi})\dot{\xi} \equiv A\frac{\dot{\xi}^3}{\dot{r}} + B_i\dot{\xi}^2 - m\frac{(\dot{r}^2 + r^2\dot{\phi}^2)}{2}, \quad i=1,2; \quad (7)$$

where $A = \tilde{p}Vr^5 e^{\frac{2V\xi}{r}}$, $B_i = -\tilde{U}_i$, $i = 1,2$. The main difference between the cases with the potential functions consists in the fact that all terms in $F_1^2$ depend on the thermodynamic velocity $\dot{r}$. For $F_2^2$ the first and third terms depend on $\dot{r}$ only. To elicit features of the generalized metric (1) we will also utilized its approximation by a NUT-metric with a NUT-parameter set and negative $\Lambda$, $\Lambda < 0$:

$$dl^2 = \Phi(R)\left(dx^0 + 4n\sin^2(\theta/2)d\phi\right)^2 - \frac{dR^2}{\Phi(R)} - \frac{R^2}{\Phi(R)}(d\theta^2 + \sin^2\theta\, d\phi^2), \quad (8)$$

and a following lapse function $\Phi(r)$ [13,14]:

$$\Phi(r) = 1 - \frac{1}{3}\Lambda(r^2 + 5n^2) + \frac{q_e^2 + q_m^2}{n^2 + r^2} - \frac{1}{n^2 + r^2}\left(rr_g - \frac{8\Lambda n^4}{3} + 2n^2\right).$$

Here $r_g = 2G^*M/c^2$, $n$ is a constant NUT-parameter, $q_e$ and $q_m$ are the electric and magnetic charges, respectively. The expression for the electrical potential may be taken as

$$\varphi = q_e\left(\frac{1}{r_+} - \frac{1}{r}\right),$$

where, $q_e/r_+$ defines the asymptotic value of the electric potential measured at infinity. A Hawking temperature $T$ and an entropy $S$ may be found from

$$S = \pi r_+^2, \quad T = \frac{1}{4\pi}\left.\frac{\partial \Phi(r)}{\partial r}\right|_{r \to r_+},$$

where outer-horizon radius of the black hole $r_+$ is given by the zeroes of the lapse function $\Phi(r)$: $\Phi(r_+) = 0$. The Hawking temperature $T$ of charged generalized-NUT black hole has the following explicit form:

$$T = \frac{1}{4\sqrt{\pi S}}\left[\frac{2S}{\pi n^2 + S}\left(-\frac{\Lambda S}{\pi} - \frac{5\Lambda n^2}{3} + 1\right) - \frac{\Lambda S}{\pi} - \frac{\pi}{\pi n^2 + S}\left(q_m^2 + 8\Lambda n^4 - 2n^2 + \frac{S\varphi^2}{\pi}\right) + \frac{5\Lambda n^2}{3} - 1\right].$$

A Gibbs free energy specific to the mixed grand-canonical-canonical ensemble involving a varying electrical charge $q_e$ and a fixed magnetic charge $q_m$ is as follows

$$G = M - TS - \varphi q_e.$$

For a charged generalized-NUT black hole the Gibbs free energy $G$ in terms of the variables $\varphi$, $q_m$ and $S$ reads

$$G = \frac{1}{4\sqrt{\pi S}}\left[-\frac{S^2}{\pi(\pi n^2 + S)}\left(\pi\Lambda n^2 - \Lambda S + 2\pi\left(-\frac{5\Lambda n^2}{3} + 1\right)\right) + S(-2\Lambda n^2 - 5\Lambda n^2 + 3)\right.$$
$$\left.+ \frac{\pi q_m^2(2\pi n^2 + 3S)}{\pi n^2 + S} + 2\pi n^2\left(-\frac{5\Lambda n^2}{3} + 1\right) - \frac{\pi(3S + 2\pi n^2)}{\pi n^2 + S}\left(\frac{8\Lambda n^4}{3} - 2n^2\right) - \frac{(S + 2\pi n^2)S\varphi^2}{\pi n^2 + S}\right].$$

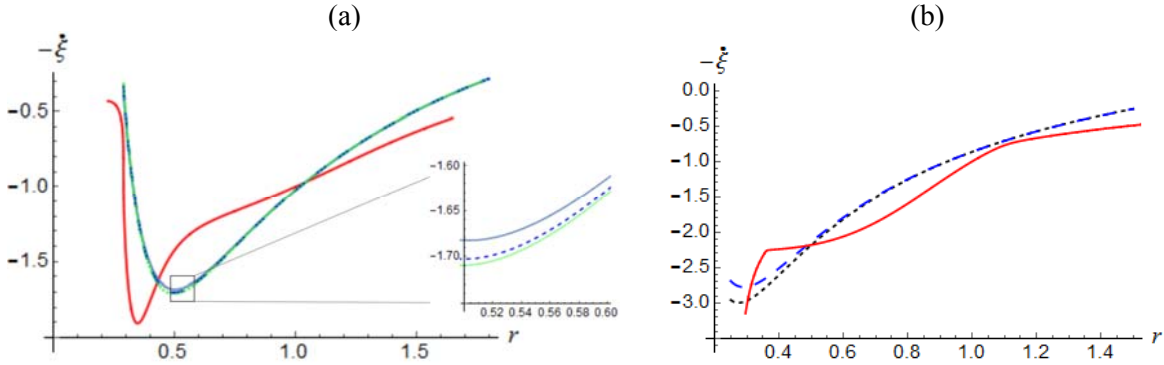

(a)          (b)

Fig. 2. The dependence $\dot{\xi}$ on $r$ along the geodesics and its approximations by lapse functions $\Phi(r)$. Thermodynamic metric-function parameters and potential functions used for numerical simulation at $V = 0.224$, $\tilde{p} = 1$: (a) $mc^2 = 1$, $m = 10^{-5}$, $U_1$; (b) $m = 1$, $c = 1$, $U_2$. Approximation parameters: (a) $\Lambda = -0.01$, $r_g = 2.7$, $q_e^2 + q_m^2 = 0.82$, $n=0.1$ (grey solid line), $n=0.05$ (blue dashed line), $n=0.01$ (green dotted line); (b) $\Lambda = -0.1$, $r_g = 2.2$, $q_e^2 + q_m^2 = 0.55$, $n=0.1$ (blue solid line), $n=0.01$ (black dashed line).

## Results

Let calculate the thermodynamic velocity $\dot{\xi}(r)$ determining the lapse function $f(r)$ and approximate it by a set of lapse functions $\Phi(r)$. Figs. 2a-b demonstrate the dependence of $\dot{\xi}(r)$ along the thermodynamic geodesics obtained by the numerical simulation for potential functions $U_1$ and $U_2$. The corresponding approximations by the generalized NUT-metric has been performed (see fig. 2). As one can see, the $\dot{\xi}(r)$ dependence for potentials $U_i$, $i = 1,2$ are in good agreement with the lapse functions $\Phi(r)$. It should be noted that the lapse functions $\Phi(r)$ is weakly dependent on the NUT-parameter $n$. For example, the curves obtained for $n=0.1$, $0.05$ and $0.01$ practically coincide, as shown in fig. 2a.

We has been studied the thermodynamics of the charged generalized NUT black hole with parameters obtained from geodesics of the metric function (3). In figs. 3 and 4 the dependencies of

Gibbs energy $G$ on the Hawking temperature $T$ and the plot of $T$ with the entropy $S$ can hold cusp points. In cusp points, there is sudden change in a physical solutions such as a pitchfork bifurcation and a spontaneous symmetry breaking in system happens [15]. For the case corresponding to the Finsler function $F_1^2$ there are two cusp bifurcations and an additional sudden-change point named "swallowtail point" (fig. 3a).

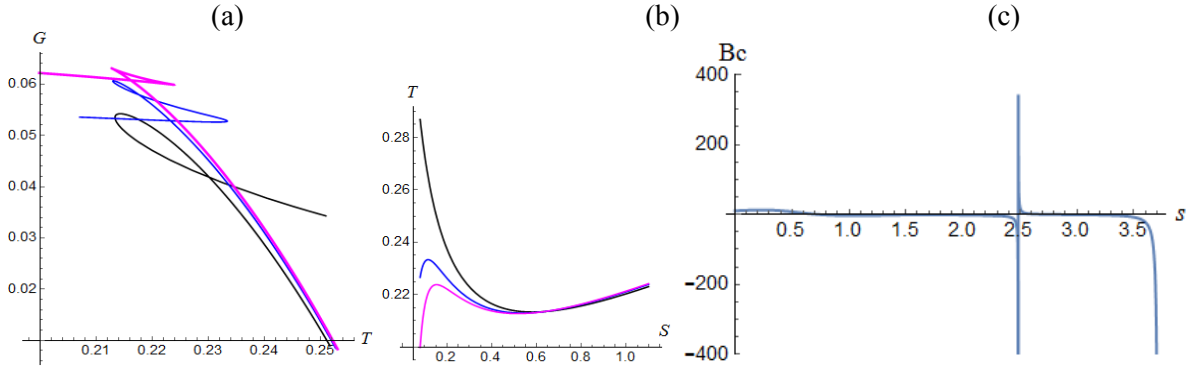

Fig. 3. The dependences of $G(T)$ (a), $T(S)$ (b) and $B_C(s)$ (c) for the metric with the potential $U_1$. The black, blue and magenta curves correspond to $n=0.1$, $n=0.05$ and $n=0.01$, respectively, at $\Lambda = -0.01$, $q_m=0.09$, $\varphi=0.6$.

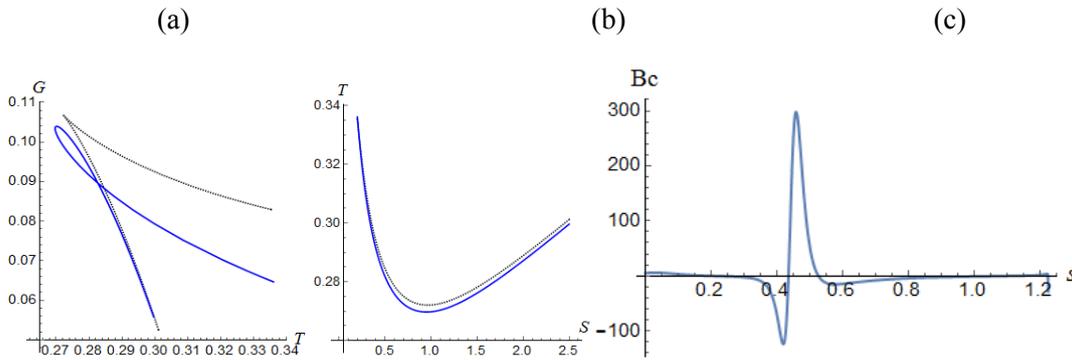

Fig. 4. The dependences of $G(T)$ (a), $T(S)$ (b) and $B_C(s)$ (c) for the metric with the the potential $U_2$. The blue and black curves correspond to $n=0.1$ and $n=0.01$, respectively, at $\Lambda = -0.1$, $q_m=0.09$, $\varphi=0.01$.

As the second order phase transitions, the first order phase transitions can be geometrically interpreted as singularities of a space curvature [16,17]. The latter implies that the curvature tends to infinity in the points where derivations of a thermodynamic potential diverge. The zero-derivations give a significant arbitrariness in a choice of the metric function.

Let analyze a behaviour of a Berwald curvature calculated along the geodesics. To do this we find the mean Berwald curvature tensor $E_{ij}$ according to the following definition:

$$E_{ij} = \frac{1}{2} \frac{\partial^2}{\partial y^i \partial y^j} \sum_k \frac{\partial G^k}{\partial y^k},$$

where $\{y^i\}_{i=1}^3 = (\dot{r}, \dot{\phi}, \dot{\xi})$, the spray coefficients $G_i$ may be expressed through the metric tensor as

$$G^i = \frac{1}{4} g^{il} \left\{ 2 \frac{\partial g_{jl}}{\partial x^k} - \frac{\partial g_{jk}}{\partial x^l} \right\} y^j y^k, \quad g_{kl} = \frac{1}{2} \frac{\partial F^2}{\partial y^k \partial y^l}.$$

A scalar Berwald curvature can be determined as the convolution product of the mean Berwald curvature tensor $E_{ij}$ with the metric tensor:

$$B_C = g^{ij} E_{ij}.$$

As one can see from figs. 3c, 4c, and 5c,f, the Berwald curvature $B_C$ changes sign in the phase transition exhibited by the isotherms $s(r)$ represented in fig. 1b and 5b,e. Increasing the number of cusp points and appearing the swallowtail point narrows the phase transition sharply but does not change its qualitative character. Because of this we will further analyze the geometrodynamics in the case of $U_2$. The lapse function $\dot{\xi}$ is nonzero in fig. 5a and trends to zero in fig. 5d in the phase-transition area. Therefore, one observes an emerging black hole because $B_C$ of the metastable "liquid phase state" is negative. After the phase transition, since $B_C$ becomes positive, the state of the black hole is stabilized (see fig. 5f).

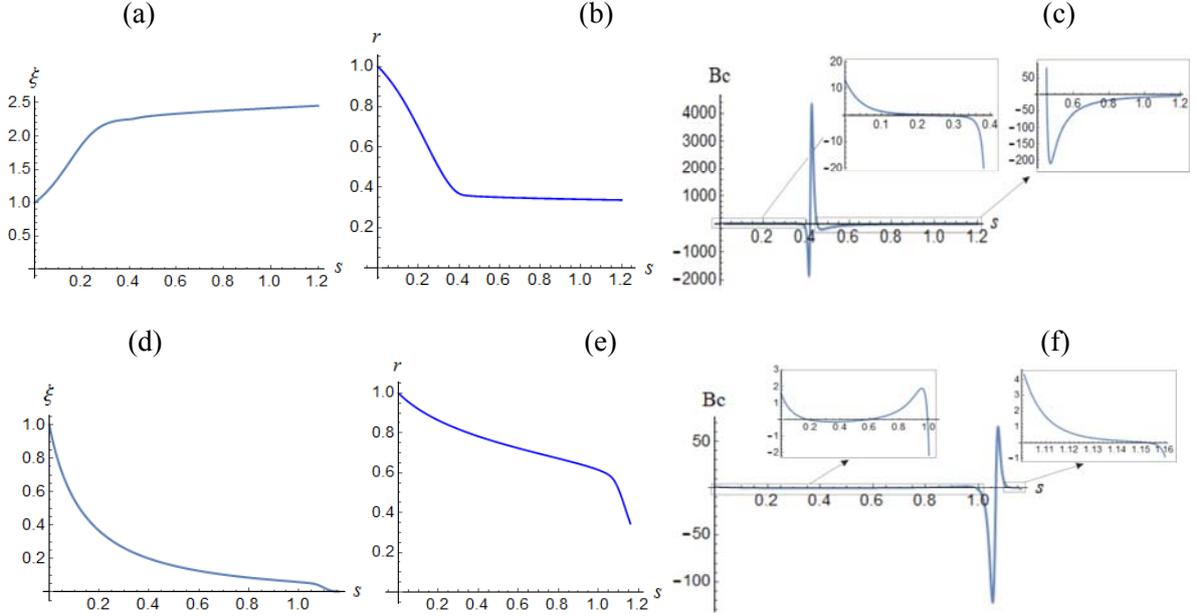

Fig. 5. The dependences $\dot{\xi}(s)$ (a, d), $r(s)$ (b, e), $B_C(s)$ (c, f) along the geodesics for the thermodynamic metric spaces with the potential $U_2$. Metric function parameters used for numerical simulation: $\tilde{p}=1$, $m=1$, $c=1$; $V=0.05$ (a-c), $V=10$ (d-f).

## Discussion and conclusion

The solutions obtained above are similarly to the ones for black hole, but these solutions are not singular as it follows from fig. 5a. Domains of emerging "liquid phase" play a role of metastable black holes. At very large compressed rates a relaxation time $\dot{\xi}(s)$ (inverse expansion coefficient) and, respectively, the lapse function $f(r)$ become zero (see figs. 2, 5d). This singularity may be interpreted as a non-zero shift in the direction of 5th dimension by the analogy with a finite value of displacement of Langmuir monolayer at its separation from a subphase surface during the 1-st order 2D phase transition. Accordingly, the non-singular black-hole-like solution describes emerging black holes with zero separation distance from the 4D spacetime as a hypersurface in 5-dimensional manifold.

So, we studied the geometrothermodynamics of electrically charged and dyonic charged black hole occurring in spacetime with generalized Newman–Unti–Tamburino (NUT) metric holding a functional NUT-parameter and negative cosmological constant. It has been shown that black-hole stability depends on sign of the curvature after the transition. We approximated the numerically calculated dependences of relaxation times on radius $r$ by the lapse functions $f(r)$ of different types and then calculated corresponding dependences of the Gibbs free energy on the Hawking temperature. We have found that bifurcations of pitchfork type as a feature of the phase transition in the NUT-black holes are revealed.